\documentclass{aa}
\usepackage{epsfig}
\begin{document}
\title {Astronomical detection of the cyanobutadiynyl radical C$_5$N}
\author{M.~Gu\'elin \inst{1} 
   \and N.~Neininger \inst{2} 
   \and J.~Cernicharo \inst{3}}
%\offprints{M.~Gu\'elin}

\institute{IRAM, 300 Rue de la piscine, F--38406 S$^{\rm t}$ Martin
           d'H\`eres, France 
      \and Radioastron. Institut der Universit\"at Bonn,
            Auf dem H\"ugel 71, D-53121 Bonn, Germany 
      \and  Instituto de Estructura de la Materia, Madrid, Spain}

\date{Received March 24, / Accepted May 5, 1998}

\thesaurus{02.13.4;08.03.4;08.16.4;09.13.2;13.19.5}

\maketitle

\markboth{Detection of C$_5$N }{Gu\'elin, Neininger, Cernicharo}

\begin{abstract}
We report the detection of the elusive carbon-chain radical C$_5$N
in the dark cloud TMC1 and its tentative detection in the circumstellar
envelope IRC+10216. C$_5$N appears to be two orders of magnitude less abundant 
than the related molecule HC$_5$N and much less abundant than expected from 
current gas phase chemistry models. In comparison the HC$_3$N to C$_3$N 
abundance ratio is of the order of 10, in reasonable agreement with model
predictions.

We have also detected in IRC+10216 two lines arising from the C$_3$H radical
in its excited $\nu_4=1$ state. 

\end{abstract}     

\section{Introduction}

That long carbon chain radicals could be abundant and play a large role
in interstellar chemistry was first recognized with the
discoveries of C$_3$N in the circumstellar envelope IRC+10216
(Gu\'elin \& Thaddeus 1977) and in the dark cloud TMC1
(Friberg et al.\ 1980). To date seven acetylenic chain
radicals, C$_n$H, $n= 2-8$,  and five cyanopolyyne 
molecules, HC$_{2n}$CN, $n=1-5$, are identified in TMC1 and/or IRC+10216 
(Bell et al.\ 1997, Gu\'elin et al.\ 1997). 
Surprisingly, no cyanopolyyne radicals heavier than C$_3$N were so far 
detected, despite model predictions that at least C$_5$N should be abundant 
(Herbst  et al.\ 1994).

The non-detection of C$_5$N was first blamed on a small dipole moment and
on the lack of spectroscopic data. The  dipole moment $\mu_0$ and the 
rotation constant $B_0$ of the  C-chains radicals depend critically 
on the nature of their electronic ground state. The two lowest states, 
$^2\Pi$ and $^2\Sigma$, 
are close in energy (Pauzat et al.\ 1991). Pauzat and co-workers
predicted from unrestricted Hartree-Fock calculations  that the 
$^2\Pi$ state  of C$_5$N lay below the $^2\Sigma$ state and that 
$\mu_0$  was  very small. The line strengths 
scaling with $\mu_0^2$,  it was no wonder that C$_5$N  escaped detection.

Things changed when Botschwina (1996) showed from more elaborate 
coupled cluster calculations that the  C$_5$N ground state was in fact 
$^2\Sigma$ and that $\mu_0$ was as large as 3.385~D, even larger than 
the dipole moment of C$_3$N. It then became clear that C$_5$N could be 
detected at least in the laboratory. Kasai et al.\ (1997) succeeded to 
synthetize this species, to measure its microwave spectrum, and to  
derive  its rotational, fine and hyperfine constants,
making at last possible a sensitive search for C$_5$N in space. 
In this Letter, we report the astronomical detection of this radical.

\section{Observations and results}

The long carbon-chain molecules and radicals are nearly thermalized in 
TMC1 and IRC+10216  with rotation temperatures, $T_{rot}$, in  the
ranges 6--10 K and 20--50 K, respectively. Since, for C$_5$N, 
$\frac{hB_0}{k}= 0.067$ K, the strongest  lines  in these two sources 
lie in the $\lambda=$ 1 cm and $3-4$ mm 
atmospherical windows. We thus searched for C$_5$N at these wavelengths, using 
the Effelsberg 100-m telescope and the Pico Veleta 30-m telescope.

The Effelsberg observations were made in October 1997, January and March 1998.
The  telescope was equipped with the K-band maser receiver. The 
weather was mostly clear and the system temperature $T_{sys} \simeq 100$ K. 
We used the new AK90 autocorrelator split into two 20\,Mhz-wide bands 
of 4048 channels each. One of the bands covered
both fine structure components of the $N=9\rightarrow 8$ transition
of C$_5$N  (25.250 GHz, Kasai et al.\ 1997), the
other the $J= 21.5\rightarrow 20.5$ transition of $^2\Pi_{3/2}$ C$_8$H
(25.227 GHz). 
The local oscillator frequency was 
switched by $\pm 200$ kHz or $\pm 500$ kHz and the spectra folded accordingly. 
The average  
spectrum obtained in TMC1 at the position of the cyanopolyyne peak
(1950.0: $\rm 4^h38^m38.6^s, 25^{\circ}35'45''$), smoothed to 15 kHz
(0.17 kms$^{-1}$), has a r.m.s. noise of 3.5 mK (units of $T_{MB}$).

In addition to C$_5$N and C$_8$H, we observed briefly the 
$J=10\rightarrow 9$ line of HC$_5$N (23.96 GHz) and the 
$N= 2\rightarrow 1$ lines of C$_3$N (19.79 GHz). 
All the data were calibrated following the procedure described by Schilke 
\& Walmsley (1991) by observing at every frequency the planetary nebula 
NGC~7027, whose flux was taken equal to 5.8 Jy. The calibration uncertainty,
which results mostly from the atmospheric absorption correction and from
beam efficiency variations, is estimated to be $< 20\%$. The HC$_5$N and C$_3$N
line intensities we measure are consistent with those reported by T\"olle et 
al.\ (1981) and Gu\'elin et al.\ (1982). 

Figure~1 (see also Table~1) shows the spectrum covering the C$_5$N transitions.
We see two 0.3 kms$^{-1}$-wide spectral lines, each detected at 
$>5 \sigma$. The lines are separated by $10.71\pm.01$ MHz, which is very 
close to the value of the spin-rotation constant measured by Kasai et 
al.\ (1997), $\gamma= 10.75$ MHz. Their half-power width is comparable to 
the C$_3$N linewidth, $0.23\pm .05$ kms$^{-1}$. Their 
rest frequencies coincide within the small uncertainties with the 
C$_5$N N=9--8 transition frequencies calculated by Kasai et al.\ (1997), 
if we adopt the source LSR velocity of $5.65\pm 0.05$ kms$^{-1}$ measured 
for C$_3$N and C$_4$H (see Gu\'elin et al.\ 1982).
Since there are no other comparable lines in the 20 MHz-wide spectrum we 
observed, the probability for a chance coincidence is
$< 10^{-6}$. We thus conclude that we have detected C$_5$N in TMC1. 

\begin{figure}
\epsfig{file=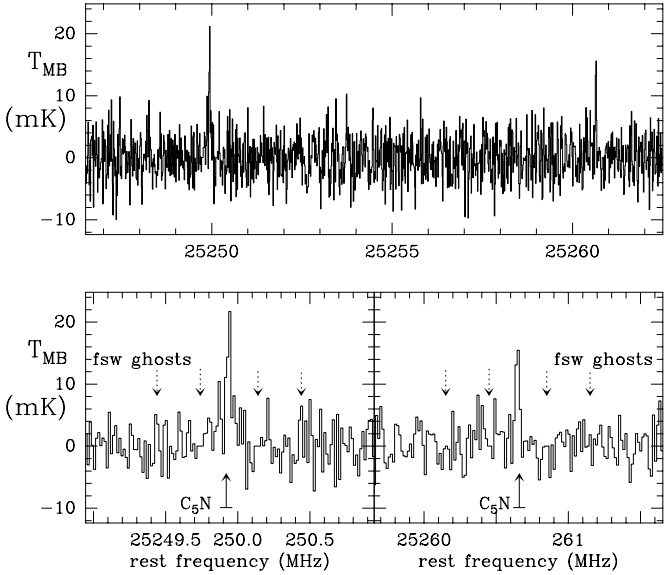,width=8.8cm,angle=0}
%\vspace{0.5cm}
\caption{Spectrum observed with the 100-m Effelsberg telescope toward
TMC~1 (cyanopolyyne peak: $\rm 4^h38^m38.6^s, 25^{\circ}35'45''$, 1950.0) 
and covering the N=9--8 rotational transition 
of C$_5$N. The N=9--8 line is split into
two fine structure components whose frequencies, derived from 
laboratory measurements, are marked by upward 
arrows. The hyperfine structure is too small to
be resolved. The spectrum was observed by switching the local 
oscillator in frequency by $\pm 200$ kHz or $\pm 500$ kHz; the 
position of the ghosts of the line in the 
folded spectrum are indicated by downward arrows.
}
\end{figure}

The 30-m observations were made in Nov.\ 1997 and April 98. The telescope 
was equipped with two  SIS mixer receivers with orthogonal polarizations. 
The zenith atmospheric opacity was below 0.1 and the system temperature 
$T_{sys}= 150-200$ K. The observations were made by wobbling 
in azimuth the secondary mirror at a rate of 0.5 Hz and with an amplitude of
90$''$. We searched  
for the N=32--31 line, the next two lower lines being partly blended with 
lines of unrelated species. The average spectrum obtained
toward IRC+10216 is plotted in Figure 2a. It has an r.m.s.\ noise per 1 MHz 
channel of 0.90 mK in the $T_A^*$ scale.

\begin{figure}
\psfig{file=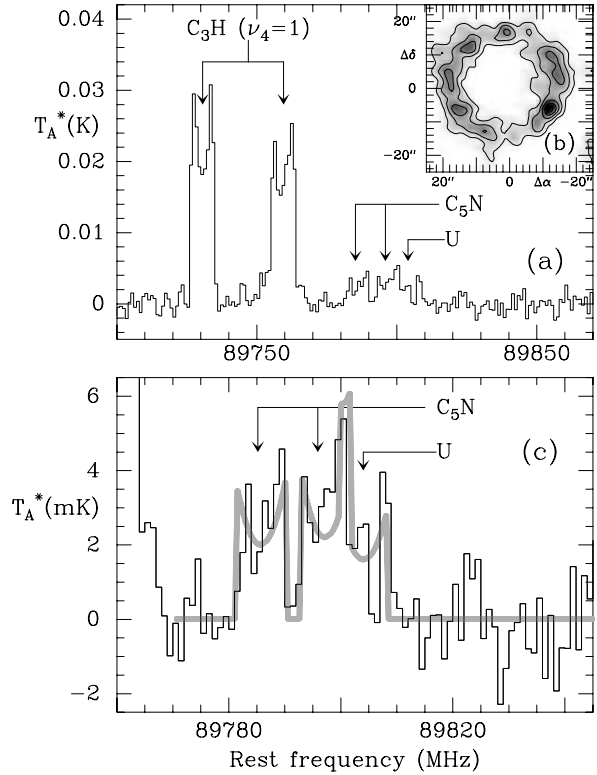,width=10cm,angle=0}
%\vspace{0.5cm}
\caption{{\bf a)}: Spectrum observed with the 30-m telescope toward IRC+10216.
The C$_3$H and C$_5$N line frequencies derived from laboratory
measurements are indicated by downward
arrows. The spectral resolution is 1 MHz (3.3 kms$^{-1}$). 
{\bf b)}: 
The J=4.5--3.5 line emission of C$_3$H in its $^2\Pi_{1/2}$
ground vibrational state, observed in IRC+10216 with the IRAM interferometer
(Gu\'elin, Lucas \& Cernicharo 1993).
 The line intensity has been integrated over a narrow velocity interval 
(2 kms$^{-1}$) centred on the star velocity. 
The coordinates represent the offsets in r.a. and dec. with respect to the 
central star. {\bf 2c)}: The central half of the spectrum of Fig.~2a, 
compared to the fitted 3-line spectrum 
(see text).
}
\end{figure}

The carbon-chain molecules observed in IRC+10216 are concentrated in a
thin shell of radius $\simeq 15''$ (see Fig.2b). Observed with the 30-m 
telescope, the profiles of their 
$\lambda$~3 mm lines have the same cusped shape and the same
width (29.5 kms$^{-1}$). They are all centred at 
$V_{\rm LSR}$= -26.5 kms$^{-1}$.
The spectrum of Fig.~2a shows two cusped lines,
near 89.75 GHz and three weaker features near 89.80 GHz. The former
can be readily assigned to the $N=4\rightarrow 3, F=4.5\rightarrow 3.5$
and $F=3.5\rightarrow 2.5$ fine structure components of C$_3$H in 
its first excited bending state, $\nu_4=1,^2\Sigma^\mu$ 
(Yamamoto et al.\ 1990)
since they have the right frequencies and the right intensity ratio 
($1.33\pm .1$, whereas the expected LTE ratio is 1.30).

We have fitted 3 cusped lines (assuming standard line shapes and 
29.5 kms$^{-1}$ 
widths) to the 3 weak spectral features of Fig.~2a. The derived
intensities and rest frequencies are given in Table 1; the fitted spectrum is 
compared in Fig.~2c to the observed spectrum.
The first two lines agree in frequency with the $(N,J)= 
(32,32.5)\rightarrow (31,31.5)$
and $(32,31.5)\rightarrow (31,30.5)$ transitions of C$_5$N and very probably 
arise from this radical. Indeed, although the $\lambda$ 3 mm spectrum of 
IRC+10216 is more crowded than the 1 cm spectrum of TMC1, there are not 
many lines that we cannot assign to the rotational transition of a known 
circumstellar molecule: in our 30-m telescope spectral survey of IRC+10216,
the density of unidentifed lines stronger than $\geq 3$ mK is only 
of 1 per 80 MHz (see Gu\'elin et al.\ 1997). The probability that two
unrelated lines lie within 1 MHz from the C$_5$N lines is thus lower than 
1 per one thousand. Whereas the first two features can be tentatively 
assigned to C$_5$N, the third, which is weaker and lies 7 MHz higher 
in frequency, remains unidentified.

\vspace{0.5mm}
%\small
\begin{table*}
{\bf Table 1: Observed line parameters}\\
\begin{tabular}{lllll}
\hline
 Rest. frequ. & Obs.-Calc.& Transition & Species&  $\int T_{MB}dv$\\
      (MHz)  & (MHz)   & $N,J\rightarrow N',J'$  & (mK.kms$^{-1}$)  \\
\hline
 TMC1\\
\hline 
19799.956  &  .005     &2,1.5,1.5-1,0.5,0.5&C$_3$N& 23 (8)\\  %units to K=4.1
19800.121  &  .000    &2,1.5,2.5-1.0.5,1.5&C$_3$N& 58 (8)\\
19780.801  &  .001     &2,2.5,2.5-1,1.5,1.5&C$_3$N& 60 (8)\\
19780.826  &  .000  &2,2.5,1.5-1,1.5,0.5&C$_3$N& 34.5 (8)\\
19781.096  &  .002 &2,2.5,5.5-1,1.5,1.5&C$_3$N& 86 (8)\\

23963.897& 0.000 & 9-8& HC$_5$N& 2240 (100)\\

25249.938 (4)  & .018 (40)&9,9.5-8,8.5& C$_5$N &7.3 (9)\\  %units to K= 6.16
25260.649 (4)  & -.017 (40)&9,8.5-8,7.5& C$_5$N &6.4 (9)\\

\hline
IRC+10216\\
\hline
89730.54(10)&-.07 (10)$^b$ & 4,4.5 - 3,3.5 &C$_3$H($\nu_4^2\Sigma^\mu$)&880(20)\\
89759.17(12)&-.18 (12)$^b$ & 4,3.5 - 3,2.5 &C$_3$H($\nu_4^2\Sigma^\mu$)&660(20)\\  
89785.6 (4) & .4 (13) &32,32.5-31,31.5& C$_5$N &95 (15)\\   %Beff/Feff=0.81
89797.0 (3) & 1.1(13)&32,31.5-31,30.5& C$_5$N &105(20)\\
89804.0 (10) & -- & --& U&75 (20)\\ 
\hline
\end{tabular}
\vspace{0.4cm}

Notes to the Table:\\
\noindent {\small
$^b$: weighted average of the two blended 
hyperfine components. 
\noindent The calculated frequencies are taken from Kasai et 
al.\ (1997) for C$_5$N, Yamamoto et al.\ (1990) for C$_3$H, and Gu\'elin et 
al.\ (1982) for C$_3$N (see also Gottlieb et al. 1983). The observed rest 
frequencies of C$_3$N and C$_5$N were derived  
assuming V$_{\rm LSR}$= 5.65 kms$^{-1}$ in TMC~1 and -26.5 kms$^{-1}$ in 
IRC+10216; that of HC$_5$N was derived assuming V$_{\rm LSR}$= 5.75 kms$^{-1}$ 
(see Gu\'elin et al.\ 1982). The number 
in parenthesis represent the r.m.s. uncertainty on the last digit. The 
uncertainties on the line integrated intensities, given in the Table, 
do not include the calibration uncertainty which is $10-20 \%$. Note that
the intensity ratio between the rotational transitions in the C$_3$H
ground state and in the $^2\Sigma^\mu$ excited bending state, hence 
presumably the population ratio between these two states, is $\simeq 10$.\\
 }
\end{table*}

\section{The abundance of C$_5$N}

In the direction of the TMC~1 cyanopolyyne peak, the rotational populations 
of HC$_5$N and HC$_3$N can be 
described by Boltzmann distributions with rotation temperatures of $7-10$ K
(see e.g.\ Takano et al.\ 1997). We adopt therefore T$_{rot}$= 8 K. The 
H$_2$ column density in this direction is N(H$_2$)= $10^{22}$ cm$^{-2}$
(Cernicharo \& Gu\'elin 1987). That of C$_5$N can be calculated from the
standard expression for the optically thin lines of thermalized linear 
molecules:

$$ {\rm N(C_5N)}= 0.7\, 10^{16}\frac{T_{rot}T'_{rot}}{T'_{rot}-T'_{bg}} 
(\nu\mu)^{-2}e^{E_u/kT_{rot}}\Sigma\int{T_{MB}dv} , $$

\noindent where $$ T'= (h\nu /k)(e^{\frac{h\nu}{kT}}-1)^{-1},$$

\noindent 
and where $\Sigma\int{T_{MB}dv}= 0.014$ Kkms$^{-1}$ is the sum of the 
integrated  intensities of the two doublet components. In equation [1],
$N$ is in cm$^{-2}$,
the dipole moment $\mu_0$ = 3.385~D (Botschwina 1996) in debye, and
the line frequency $\nu= 25.25$ GHz in gigahertz. We find:

\noindent $ {\rm N(C_5N)}= 3.1\, 10^{11}\,{\rm cm^{-2} \, ,}$\\ 
$x({\rm C_5N)=N(C_5N)/N(H_2)}\simeq 3\, 10^{-11}$.
 
The abundance of C$_5$N can be compared to those of the related species 
HC$_5$N and C$_3$N in the light of the chemical model predictions.
We calculate first the abundance of the $^{13}$C isotopomers of HC$_5$N,
whose 1 cm lines are optically thin, by setting in equation [1] \,
$\mu_0= 4.33$ D, $\nu= 23.7$ GHz and  $\int{T_{MB}dv}= 
0.065$ Kkms$^{-1}$, which is the average of the integrated intensities of 
the J=9--8 lines of HC$^{13}$CC$_3$N and HC$_4\,$$^{13}$CN, observed  
with the Effelsberg telescope (Takano et al. 1998). We find 
N(HC$_4\,$$^{13}$CN)= 
1.0 $10^{12}$ cm$^{-2}$, from which we derive N(HC$_5$N)= 7 10$^{13}$ 
cm$^{-2}$, adopting the `standard' elemental abundance ratio 
$^{12}$C/$^{13}$C$\simeq$70 in the local interstellar medium 
(see Wilson \& Rood 1994). The value of 7 10$^{13}$cm$^{-2}$ is close
to the value of Suzuki et al.\ (1992), as well as to the value
we estimate with an LVG code from the main isotopomer 
$J=9\rightarrow 8$ line intensity (5 10$^{13}$cm$^{-2}$). 
We arrive at:

\noindent N(HC$_4\,$$^{13}$CN/C$_5$N)= 3 \\
N(HC$_5$N/C$_5$N)$\simeq 200$.

For C$_3$N, we find, using in eq. [1] $T_{rot}=8$K, $\mu_0=2.84$ D
(Pauzat et al. 1991), and the line parameters 
of Table 1: N(C$_3$N)= 8.2 10$^{12}$ 
cm$^{-2}$. Finally, we take for HC$_3$N
the column densities derived by 
Takano et al.\ (1998), \\
N(HC$^{13}$CCN)=
2.1 10$^{12}$ cm$^{-2}$  and 
N(H$^{12}$C$_3$N)= 1.6 10$^{14}$ cm$^{-2}$. This yields: 
 
\noindent N(HC$^{13}$CCN/C$_3$N)= 0.26 \\
N(H$^{12}$C$_3$N/C$_3$N)= 19.

\noindent The C$_5$N/HC$_5$N abundance ratio is an order of magnitude smaller
than the C$_3$N/HC$_3$N ratio.  According to neutral-neutral gas phase 
chemical models, the cyanopolyyne mole-cules are mainly formed by 
reactions of N and CN with polyacetylenes or polyacetylic ions
(Herbst \& Leung 1990). The  cyanopolyyne radicals are formed in 
TMC1 by the reaction of atomic C with cyanopolyynes 
(Herbst et al.\ 1994),  and in IRC+10216 by 
photodissociation (Cherchneff \& Glassgold 1993). They are destroyed
by reactions with N atoms, O atoms, polyacetylenes and photodissociation. 

According to model predictions, HC$_3$N/HC$_5$N varies in TMC1 by several 
orders of magnitude between the ``early times'' and steady state. The 
C$_3$N/HC$_3$N  and C$_5$N/HC$_5$N ratios, on the other hand, remain constant 
within a factor of 2. They are comprised between 0.1 and 0.2
(see Table 3 of Herbst et al. 1994). Whereas the observed C$_3$N/HC$_3$N ratio
(0.2, Cernicharo et al.\ 1987) agrees with the predicted one,
C$_5$N/HC$_5$N is more than one order of magnitude too low. The most recent 
models, which take into account the destruction of acetylene and polyacetylenes 
by C atoms and the reaction of CN  with O, form enough
C$_5$N but too little C$_3$N, HC$_3$N and HC$_5$N (Herbst et al.\ 1994).  

In the case of IRC+10216, we adopt for C$_5$N the same rotation temperature 
as for HC$_5$N ($T_{rot}$= 29 K, 
Kawaguchi et al.\ 1996). We then derive from the integrated intensities 
of Table 1 a line-of-sight column density in the direction of the central star 
(twice the radial column density across the shell) 
N(C$_5$N)$=6\, 10^{12}$ cm$^{-2}$. This is $\sim 50$ times less than 
the column densities of C$_3$N and HC$_5$N. 
Here also,  C$_5$N is underabundant with respect to model predictions
(Cherchneff \& Glassgold 1993), and C$_5$N/HC$_5$N one order of magnitude 
smaller than  C$_3$N/HC$_3$N.

The unexpectedly low C$_5$N/HC$_5$N abundance ratio in found both sources 
shows that the formation of long carbon-chain molecules 
is not fully understood, and that is difficult to 
predict the abundances of unobserved 
species. The very long carbon chains could be
more abundant than expected. We note, however, that the species
of medium size, such as the chains consisting of $4-7$ C,N, or O
atoms and the rings with less than 10 heavy atoms, which would give rise in 
TMC~1 to a rich centimetric spectrum, are probably
not very abundant in that source: we have covered so far 
a 100 MHz-wide band in TMC~1 with a very good sensitivity and did not 
detect any unidentified line down to a level of 10~mK. The acetylenic 
chains and cumulene carbenes appear in this respect exceptional.  

\begin{acknowledgements}
We thank W. and H. Wiedenh\"over of the MPIfR who built the new autocorrelator 
and made it available to us, W. Zinz who helped us to configurate it
for our observing runs, Dr.\ P. Schilke for advice on data
calibration, and the referee for helpful comments. 
\end{acknowledgements}

 \section{References}
Bell M.B. et al.\ 1997, ApJ. 483, L61\\
Botschwina, P. 1996, Chem. Phys. Lett. 259, 627\\
Cernicharo, J., Gu\'elin, M. 1987, A\&A 176, 299\\
Cernicharo, J. Gu\'elin, M., Menten, K.M., Walmsley, C.M. 1987, A\&A 181, L1\\
Cherchneff, I., Glassgold, A.E. 1993, ApJ 419, L41\\
Friberg, P., Hjalmarson, A., Irvine, W.M., Gu\'elin, M. 1980, ApJ 241, L99\\
Gottlieb, C.A., Gottlieb, E.W., Thaddeus, P., Kawamura, H. 1983,
ApJ. 275, 916\\
Gu\'elin, M., Thaddeus, P. 1977, ApJ 212, L81\\
Gu\'elin, M., Friberg, P., Mezaoui A. 1982, A\&A 109, 23\\
Gu\'elin, M., Lucas, R., Cernicharo, J. 1993, A\&A 280, L19\\
Gu\'elin et al.\ 1997, A\&A 317, L1\\
Herbst, E., Leung, C.M. 1990, A\&A 233, 177\\
Herbst, E., Ho-Hsin Lee, Howe, D.A., Millar, T.J. 1994, MNRAS 268, 335\\
Kasai, Y., Sumiyoshi, Y., Endo, Y., Kawaguchi, K. 1997, ApJ, 477, L65\\
Kawaguchi, K,, Kasai, Y., Ishikawa, S., Kaifu, N. 1995, PASJ 47, 853\\ 
Pauzat, F. , Ellinger, Y., McLean, A.D. 1991, ApJ 369, L13\\
Schilke, P., Walmsley, C.M. 1991, MPIfR {\it Technischer Bericht Nr.\ 70, 
July 1991}\\
Suzuki, H. et al.\ 1992, ApJ 392, 551\\
Takano, S. et al.\ 1998, A\&A {\it in press}\\
T\"olle et al.\ 1981, A\&A 95, 143\\
Wilson, T.L., Rood, R.T. 1994, ARA\&A 32, 191\\
Yamamoto, S., Saito, S., Ohishi, M. 1990, ApJ 348, 363.\\

\end{document}